\documentclass[12pt, onecolumn, letterpaper, journal, draftcls, peerreview, doublespace]{IEEEtran}

\usepackage{amssymb,amsmath,epsfig,tabularx,delarray,enumerate,subfigure,graphicx,array,cite,mathdots,color}

\providecommand{\keywords}[1]{\textbf{\textit{Index terms---}} #1}

\begin{document}
\title{Molecular Communications at the Macroscale: A Novel Framework for Modeling Epidemic Spreading and Mitigation}
\author{Yifan Chen, Yu Zhou, Ross Murch, and Tadashi Nakano
\thanks{Y. Chen is with the Faculty of Science and Engineering and the Faculty
of Computing and Mathematical Sciences, the University of Waikato,
Hamilton, New Zealand (e-mail: yifan.chen@waikato.ac.nz). He is also
with the Department of Electrical and Electronic Engineering,
Southern University of Science and Technology, Shenzhen,
China.}\thanks{Y. Zhou is with  the Department of Electrical and
Electronic Engineering, Southern University of Science and
Technology, Shenzhen, China. He is also with the Department of
Electronic and Computer Engineering, Hong Kong University of Science
and Technology, Hong Kong, China.} \thanks{R. Murch is with the
Department of Electronic and Computer Engineering, Hong Kong
University of Science and Technology, Hong Kong, China.}\thanks{T.
Nakano is with the Graduate School of Frontier Biosciences, Osaka
University, Osaka, Japan.}}

\markboth{Submitted to IEEE Transactions on Communications}{Shell
\MakeLowercase{\textit{et al.}}: Molecular Communications at the
Macroscale: A Novel Framework for Modeling Epidemic Spreading and
Mitigation}

\maketitle

\begin{abstract}
Using the notion of effective distance proposed by Brockmann and
Helbing, complex spatiotemporal processes of epidemic spreading can
be reduced to circular wave propagation patterns with well-defined
wavefronts. This hidden homogeneity of contagion phenomena enables
the mapping of \emph{virtual} mobility networks to \emph{physical}
propagation channels. Subsequently, we propose a novel
communications-inspired model of epidemic spreading and mitigation
by establishing the one-to-one correspondence between the essential
components comprising information and disease transmissions. The
epidemic processes can be regarded as \emph{macroscale} molecular
communications, in which individuals are macroscale information
molecules carrying messages (epidemiological states). We then
present the notions of normalized ensemble-average prevalence (NEAP)
and prevalence delay profile (PDP) to characterize the relative
impact and time difference of all the spreading paths, which are
analogous to the classical description methods of path loss and
power delay profile in communications. Furthermore, we introduce the
metric of root mean square (RMS) delay spread to measure the
distortion of early contagion dynamics caused by multiple infection
transmission routes. In addition, we show how social and medical
interventions can be understood from the perspectives of various
communication modules. The proposed framework provides an intuitive,
coherent, and efficient approach for characterization of the disease
outbreaks by applying the deep-rooted communications theories as the
analytical lens.
\end{abstract}


\section{Introduction}
The transmission of emergent infectious diseases has caused global
public health emergencies \cite{FRA09,CBB07,PGP14,IMG14}. The
worldwide trend of urbanization and rapid growth in connectivity
among metropolitan cities have greatly increased the risk that
highly contagious pathogens will spread. In particular,
transportation networks made of air traffic and high-speed rail have
resulted in fast and frequent fluxes of individuals. Furthermore,
the scale-free nature of the underlying mobility networks
\cite{BA99} implies that any virulent disease can spread through the
global population \cite{PV01,IKB17}. On one hand, it is critical to
implement timely and effective intervention strategies to reduce the
socioeconomic impact of a pandemic event \cite{CBB07}. On the other
hand, it becomes increasingly difficult to contain and mitigate an
outbreak \cite{CBB06,HFA07}. A key challenge is thus to design
reliable models that capture the fundamental transmission
characteristics of contagion phenomena, gain insights into the
optimal intervention strategies such as vaccination and quarantine,
and generate accurate epidemic forecasts for public health
decision-making \cite{CSB16,PCM15}.

When significant uncertainty clouds the epidemiology of an
infectious disease, phenomenological models provide a starting point
for generating early epidemic growth profiles \cite{CHO16,VSC16}.
Nevertheless, a phenomenological approach cannot evaluate which
mechanisms (e.g., airborne vs. close contact transmission model,
population behaviour changes, individual heterogeneity in
susceptibility and infectivity) might be responsible for the
empirical patterns. In this regard, a series of large-scale,
parameter-rich epidemic simulators based on mechanic transmission
models have been proposed \cite{CSB16}. Models range from
metapopulation formulations that include discrete and identifiable
subpopulations \cite{HBG04,CPV07}, spatially continuous formulations
that consider a population being continuously distributed across
space \cite{SAT09,KR08}, and individual-level formulations that
summarize disease-causing interactions at the individual scale
\cite{PCM15}.

Recently, a simple yet efficient measure derived from the underlying
mobility network has been proposed to estimate the infection arrival
time, which is given by a logarithmic \emph{ad hoc} edge weight
transformation requiring that adding edges should translate to
multiplying the associated probabilities \cite{BH13,IKB17}. This
measure is called effective distance and follows the intuitive idea
that a smaller number of passengers effectively increase the
distance between neighboring nodes. The epidemic arrival time
obtained from numerical simulations correlates highly with the
shortest-path effective distance \cite{BH13}. The correlation
becomes even higher when all walks that connect source and target
are included \cite{IKB17}. This phenomenon results in a circular
wave propagating outwards at a constant speed from the infected node
at time zero to all other nodes and permits effective detection of
the initial outbreak location. This hidden homogeneity of contagion
media permits the mapping of \emph{virtual} mobility networks to
\emph{physical} propagation channels. Nevertheless, this approach
only focuses on the statistics of the earliest arrival time of
infection \cite{GBB07,GBB08} without looking into the role and
relative importance of the multiple transmission paths in terms of
both delay and prevalence. Furthermore, the analytical framework in
\cite{BH13,IKB17} does not provide an intuitive, coherent, and
efficient approach that captures the interrelationship between the
dynamics of the initial-stage, small community-level outbreaks and
the air-traffic-mediated large-scale spreading. Finally, the model
in \cite{BH13,IKB17} does not incorporate inherently the influence
of therapeutic and social resources to minimize the impact of the
outbreak.

Here we propose a novel communications-inspired model of contagion
phenomena that remedies the aforementioned drawbacks. The initial
stage of an outbreak is the information-bearing signal, the
departure and arrival of an infected person from the origin city to
the target city by air transportation correspond to the transmission
and reception of signals, and the multiple paths linking the origin
to the target are the communication channel. Moreover, the medical
interventions such as socially targeted antiviral prophylaxis,
dynamic mass vaccination, and the social distancing at the local
scale \cite{GKL06} are modeled as pulse shaping processes, whereas
the social distancing at longer scales such as long-range air travel
restriction is modeled by the organizational flow of message
carriers (e.g., the magnetic gradient guiding the motion of message
carriers in externally controllable and trackable communications
\cite{CKA15}). The infection spreading can be regarded as
\emph{macroscale} molecular communications \cite{NAK17,NEH13,NSO14},
in which individuals are macroscale information molecules carrying
messages (epidemiological states). The proposed approach provides
new insight into disease transmission by utilizing description
methods traditionally only used in the communications field (e.g.,
path loss, delay profile, delay spread) for characterization of
epidemic growth and mitigation.

The paper is organized as follows. Section II presents the
one-to-one correspondence between the communications system and the
epidemic spreading and control process. Section III discusses the
communications-inspired description methods for epidemic spreading.
Section IV provides numerical examples to demonstrate the main
properties of the proposed modeling framework. Finally, some
concluding remarks are drawn in Section V.

\section{Analogy between Communications System and Epidemic Spreading and Control}
Fig. 1(a) shows the block diagram of a general communication system,
in which different functional elements are represented by blocks.
The analogy between the communication system and the epidemic
spreading and mitigation is also demonstrated in Fig. 1(a).

The information source produces required messages to be transmitted.
From the contagion perspective, we can consider the emergence and
non-emergence of a new index case as two opposite messages to be
broadcasted from the origin, which can be denoted by the two symbols
$``1"$ and $``0"$ in the binary code.

The modulator processes the message into a form suitable for
transmission over the communication channel. For the on-off keying
(OOK) modulation, the appearance of a message carrier for a specific
duration (e.g., a time-limited waveform) represents a binary $``1"$,
while its disappearance for the same duration represents a binary
$``0"$. In terms of disease outbreak, we can consider the initial
seeding stage of infection in a metapopulation as the modulation
process. \begin{figure} [!htp]
\begin{center}
\epsfig{file=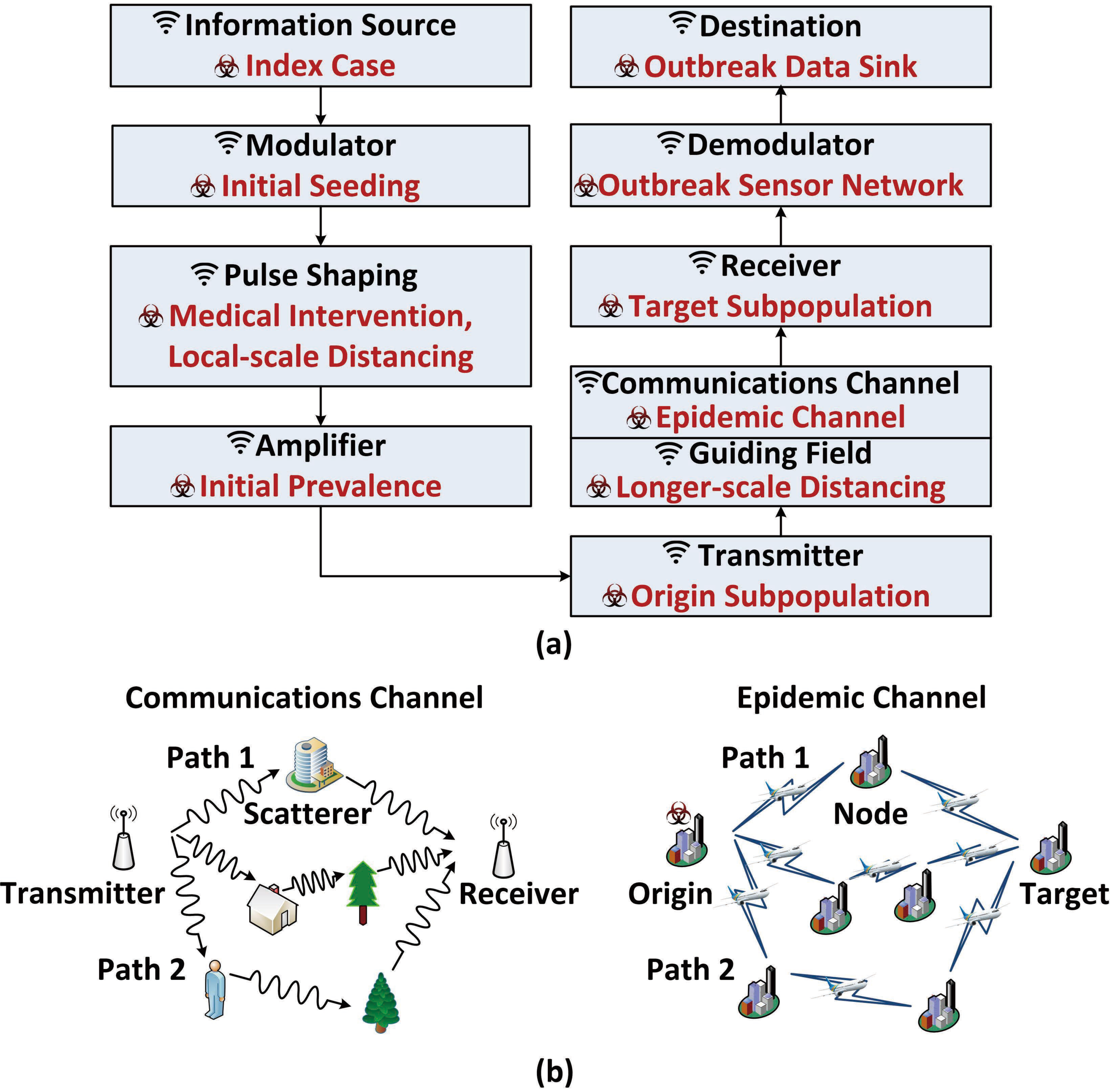,width=0.8\linewidth}
\end{center}
    \caption{(a) The block diagram showing the analogy between a general
    communications system and the epidemic spreading/mitigation process; (b) the
    one-to-one correspondence between the multipath communications channel and
    the epidemic channel.}
    \label{fig:Fig1}
\end{figure}It is supposed that the number of
susceptible individuals is much larger than the numbers of
infectious and recovered individuals, which is the common assumption
made for derivation of the effective distance and is realistic for
the usual diseases \cite{GBB07,GBB08,BH13,IKB17}. The initial
seeding is analogous to the OOK, where a burst of case incidence
$C(t)$ describing the cumulative number of cases at time $t$
corresponds to the message carrier. The presence and absence of
$C(t)$ indicate the onset (``1'') and non-existence (``0'') of a
novel pathogen, respectively. It is worth noting that ``carrier''
here is a communications terminology (i.e., to convey a message
across the medium), which should be distinguished from ``carrier''
in the medical context (i.e., a bearer and transmitter of a
causative agent of an infectious disease). For exponential growth
dynamics which has been commonly assumed for the initial growth
phase \cite{AM91}, $C(t)$ in the absence of interventions or
behavior changes is given by
\begin{equation} 
\begin{split}
C(t)=\underset{I(t)}{\underbrace{\alpha e^{\alpha
t}\theta(t)}}\otimes \underset{R(t)}{\underbrace{e^{-\beta
t}\theta(t)}},
\end{split}
\end{equation}
where $I$ and $R$ denote the infecting and recovering processes,
respectively, and $\theta(\cdot)$ is the Heaviside step function
because both $I$ and $R$ are supported on $[0,\infty]$. The term
$\alpha$ is a positive parameter denoting the growth rate and
$\beta$ is the recovery rate.

Pulse shaping is the process of changing the waveform of transmitted
signals, which can be used to describe the following intervention
strategies as illustrated in Fig. 1(a). For socially targeted
antiviral prophylaxis, symptomatic individuals and their close
contacts receive treatment or prophylaxis with antiviral drugs,
which increase the recovery rate and can thus be regarded as tuning
of $\beta$ in (1); dynamic mass vaccination and local-scale social
distancing such as school closure, voluntary behaviour change, and
formally imposed quarantine reduce the per-case growth rate for the
infecting process, which effectively reduces the parameter $\alpha$
and results in slower-than-exponential epidemic growth as observed
across a range of pathogens based on empirical data \cite{VSC16}. We
can incorporate a time-dependent $\alpha(t)$ in the model, e.g., an
exponential decline as suggested in \cite{VSC16},
$\alpha(t)=\alpha_0\left[(1-\kappa)e^{-qt}+\kappa\right]$. The rate
$\alpha(t)$ decreases exponentially from an initial value $\alpha_0$
towards $\kappa\alpha_0$ at the rate $q>0$. By substituting
$\alpha(t)$ into (1), the overall case incidence in the origin
subpopulation becomes
\begin{equation} 
\begin{split}
C(t)&=\int^t_0\alpha_0\left[(1-\kappa)e^{-q\tau}+\kappa\right]e^{\alpha_0\left[(1-\kappa)e^{-q\tau}+\kappa\right]\tau-\beta(t-\tau)}\mathrm{d}\tau\\
&\times\left[\theta(t)-\theta(t-T_\mathrm{b})\right],
\end{split}
\end{equation}
The parameter $T_\mathrm{b}$ is given by the time when the case
incidence is equal to a sufficiently small threshold $\epsilon$,
below which the outbreak is deemed quenched at the small community
level. A natural selection of $\epsilon$ is 1, which means that the
outbreak is supposed to be contained if there is only one case left.
In addition, the initial prevalence of the outbreak plays the role
of a power amplifier multiplying the transmitted pulse as shown in
Fig. 1(a).

In a multipath communications channel depicted in Fig. 1(b), the
signal propagates from the transmitter to the receiver \emph{via}
multiple scatterers, which may involve single-bounce reradiation
(e.g., Path 1) or multiple-bounce reradiation (e.g., Path 2)
\cite{MOL10}. The power density of the signal reduces as it
propagates through space. This phenomenon is called path loss or
path attenuation. Furthermore, the signal waveform may be distorted
upon reflection at a scatterer, which is depending on the frequency
response of the scatterer. Note that the aim of lossless
transmission in communications refers to the error-free delivery of
messages (i.e., ``0''s and ``1''s) rather than the message carrier
(i.e., $C(t)$); correct messages can be recovered from distorted
message carriers by using properly designed detectors at the
receiver. The arrival time introduced by each path is proportional
to its propagation length. The communications channel exhibits
structural likeness with the air-traffic-mediated epidemic channel
as also illustrated in Fig. 1(b). Multiple intermediate nodes
representing different subpopulations (cities) in the global
mobility network \cite{BH13,IKB17,GBB08,GBB07} are analogous to the
scatterers in the communications channel. A contagious disease
spreads from the origin to the target \emph{via} multiple routes due
to passenger flows on the transportation network, which have
coherent wavefronts for all paths and may involve a two-hop link
(e.g., Path 1) or a multiple-hop link (e.g., Path 2). The
probability that a successive node is contaminated by an infected
traveler from the upstream node reduces as the effective distance
between these two nodes increases as to be discussed in Section III,
which is analogous to the concept of path loss. Subsequently, the
spreading process within a subpopulation is triggered by any newly
arrived infectious individuals from elsewhere, and then forms the
local epidemiological state. This scenario is similar to the process
that a scatterer receives an incoming signal and reradiates it with
localized waveform distortion. As such, $C(t)$ in (2) can be
regarded as an ``epidemic pulse'' sent from the origin, which
``excites'' the mobility network resulting in an extended duration
of the pandemic event in the target due to multiple infection
spreading paths. This is similar to signal pulse excitation of a
multipath channel in the context of wireless communications. The
epidemic pulse starts from the origin and then propagates to all the
intermediate nodes and the target.

Finally, the demodulator is associated with a sensor network placed
in the target node to detect whether any person is infected
\cite{BH13}. Following \cite{BH13}, the locations graph is
scale-free, which allows highly efficient outbreak detection by
placing sensors in the hubs of the locations network. The
destination is the data fusion center deciding whether there is a
new primary case.

\section{Communications-inspired Description Methods for Epidemic Spreading}
For description of the epidemic channel, we start with the concept
of effective distance from a node $n$ to a connected node $m$,
$d_{m,n}=1-\ln P_{m,n}$ where the flux fraction $P_{m,n}$ represents
the fraction of travelers that leave $n$ and arrive at $m$ per unit
time and depends only on the topological features of the static
underlying network \cite{BH13,IKB17}. As demonstrated in
\cite{BH13,IKB17}, the mean arrival time is approximately
proportional to the effective distance and the effective spreading
speed depends on the epidemiological parameters, which is a global
property independent on the mobility network and the outbreak
location. Specifically, it was shown in \cite{GBB08,GBB07} that the
propagation time between two connected nodes $n$ and $m$ follows a
truncated Gumbel probability density function (pdf)
\begin{equation} 
f_{m,n}(t)=e^{1-d_{m,n}+\lambda
t-\frac{1}{\lambda}e^{1-d_{m,n}+\lambda t}}\theta(t),
\end{equation}
with average $\left\langle
t_{m,n}\right\rangle=d_{m,n}/\lambda-\left(1-\ln\lambda+\gamma\right)/\lambda$,
where $\lambda$ is the spreading rate and $\gamma$ is the Euler
constant. The Gumbel type of the arrival time distribution is
derived based on the assumption that the time to increase the number
of infectious individuals in node $n$ by one is small with respect
to the time scale of the epidemic arrival in node $m$
\cite{GBB08,GBB07,IKB17}. Hence, $\lambda$ could be approximated as
the mean value of the per-case growth rate for (2),
$\left[\mathrm{d}C(t)/\mathrm{d}t\right]/C(t)$, without much
influence on the validity of the Gumbel pdf. This can also be
verified indirectly by the empirical data of the worldwide 2009 H1N1
influenza pandemic and 2003 SARS epidemic \cite{BH13}. It was shown
in \cite{BH13} that there is a strong correlation between the
disease arrival time and the effective distance throughout the
entire time course of each disease, regardless of any mitigation
strategies implemented that may change the per-case growth rate over
time. This linear relationship is a typical feature of the Gumbel
distribution (i.e., $\left\langle t_{m,n}\right\rangle\propto
d_{m,n}$).

Next, node $m$ will only be contaminated if the arrival of an
infectious individual occurs before quenching of the outbreak in
node $n$. Hence, the probability that node $m$ is contaminated is
given by
\begin{equation} 
\rho_{m,n}\left(d_{m,n}\right)=\int^{T_\mathrm{b}}_0f_{m,n}(t)\mathrm{d}t,
\end{equation}
The quantity $\rho_{m,n}$ essentially measures the \emph{normalized}
ensemble-average prevalence (NEAP) of contagion in node $m$.
Subsequently, for an ordered path from node $1$ to node $L$,
$\Lambda=\left\{1,2,\cdots,L\right\}$, the NEAP is derived as
\begin{equation} 
\rho_\Lambda=\rho_{2,1}\left(d_{2,1}\right)\times\rho_{3,2}\left(d_{3,2}\right)\times\cdots\rho_{L,L-1}\left(d_{L,L-1}\right).
\end{equation}
This expression is analogous to the path loss in a communications
channel where the overall loss is equal to the multiplication of
attenuation for each reradiation, which is a function of the
distance of each hop. In addition, any arrival time greater than the
duration of the transmitted epidemic waveform $T_\mathrm{b}$ would
not have pathogenical significance because the passenger traveling
from node $n$ to node $m$ will not cause disease transmission after
the outbreak is extinguished at $n$. Hence, the pdf of effective
propagation time is a further truncated Gumbel distribution:
\begin{equation} 
\hat{f}_{m,n}(\hat{t})=\mu e^{1-d_{m,n}+\lambda
\hat{t}-\frac{1}{\lambda}e^{1-d_{m,n}+\lambda
\hat{t}}}\left[\theta(\hat{t})-\theta\left(\hat{t}-T_\mathrm{b}\right)\right],
\end{equation}
with $\mu$ being a normalization factor to ensure that
$\hat{f}_{m,n}(\hat{t})$ is a pdf. The mean value can be derived
after some manipulations as
\begin{equation} 
\begin{split}
\left\langle\hat{t}_{m,n}\right\rangle&=
\frac{\mu}{\lambda}\Bigg[\frac{\partial\gamma\left(s,e^{1-d_{m,n}+\lambda
T_\mathrm{b}-\ln\lambda}\right)}{\partial
s}-\frac{\partial\gamma\left(s,e^{1-d_{m,n}-\ln\lambda}\right)}{\partial
s}\Bigg]\Bigg|_{s=1}\\
&+\frac{1-d_{m,n}-\ln\lambda}{\lambda}\left(e^{-e^{1-d_{m,n}+\lambda
T_\mathrm{b}-\ln\lambda}}-e^{-e^{1-d_{m,n}-\ln\lambda}}\right),
\end{split}
\end{equation}\begin{figure} [!htp]
\begin{center}
\epsfig{file=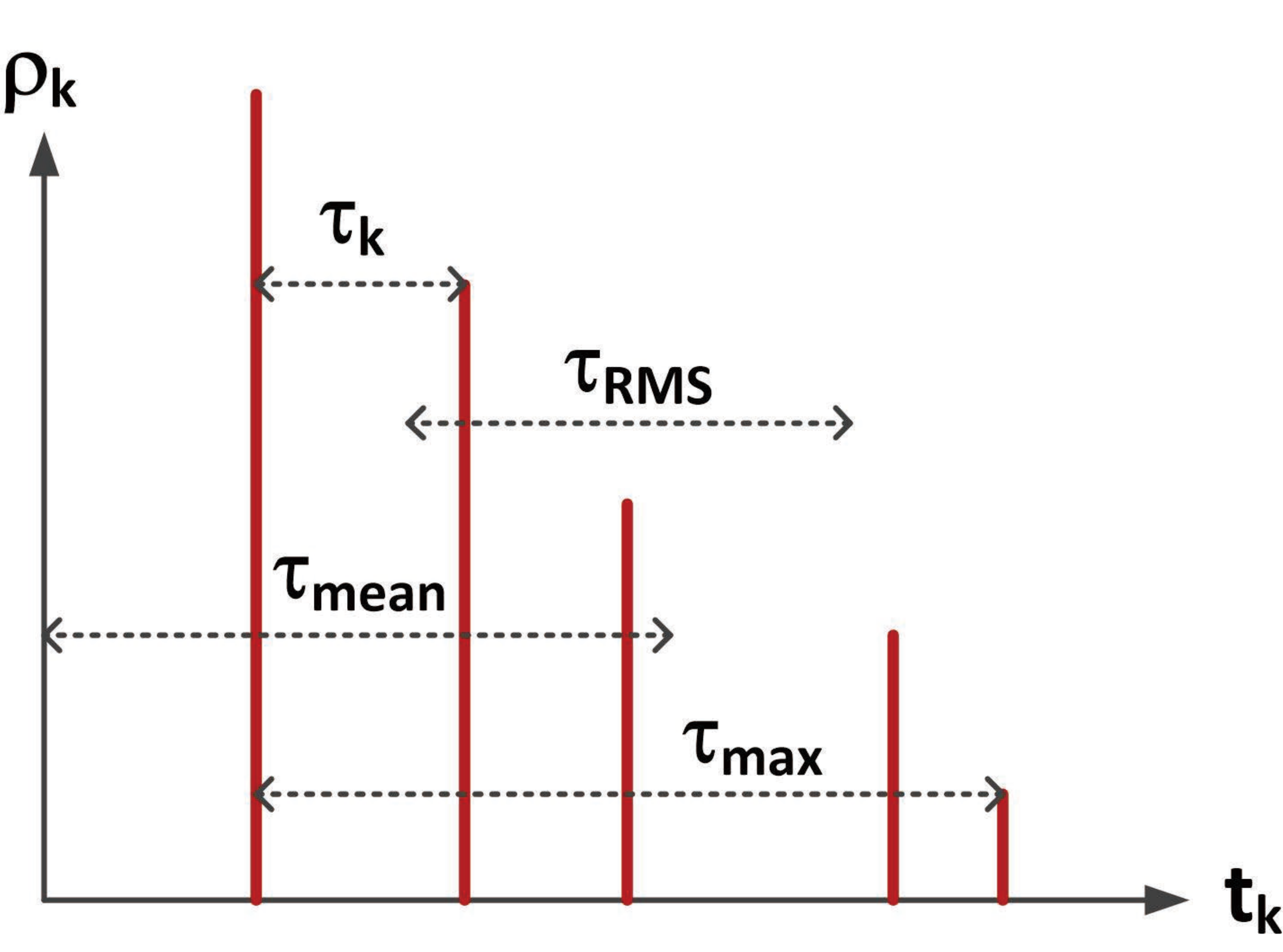,width=0.5\linewidth}
\end{center}
    \caption{Key parameters of the PDP.}
    \label{fig:Fig2}
\end{figure}
where $\gamma\left(s,x\right)$ is the lower incomplete gamma
function. Note that
$\left\langle\hat{t}_{m,n}\right\rangle\approx\left\langle
t_{m,n}\right\rangle$ if $T_\mathrm{b}\gg\left\langle
t_{m,n}\right\rangle$. For the ordered path
$\Lambda=\left\{1,2,\cdots,L\right\}$, the average arrival time is
given by
\begin{equation} 
t_\Lambda=\left\langle\hat{t}_{2,1}\right\rangle+\left\langle\hat{t}_{3,2}\right\rangle+\cdots+\left\langle\hat{t}_{L,L-1}\right\rangle.
\end{equation}

Consider a multipath epidemic channel as shown in Fig. 1(b), the
$k^\mathrm{th}~(k=1,2,\cdots,K)$ path introduces a mean delay $t_k$
and a NEAP $\rho_k$ obtained from (8) and (5), respectively.
Analogous to the classical power delay profile in communications
\cite{MOL10}, we introduce the description method of
\emph{prevalence delay profile} (PDP) as follows
\begin{equation} 
\begin{split}
\mathbb{P}(t)=\sum^K_{k=1}\rho_k\delta\left(t-t_k\right),~~~t_1\leq
t_2\leq\cdots\leq t_K,
\end{split}
\end{equation}
where $\delta(\cdot)$ is the Dirac delta function. The PDP will give
us a clear understanding of the roles of different disease
transmission routes, which is of primary importance for the study of
path-oriented contagion phenomena and the setup of containment
measures and policies.

We continue to define the following parameters derived from the PDP
as illustrated in Fig. 2, which are analogous to their counterpart
quantities in wireless communications:
\begin{itemize}
    \item \emph{Excess delay:} The delay of each path relative to
    the first arriving path, $\tau_k=t_k-t_1~(k=1,2,\cdots,K)$.
    \item \emph{Total excess delay:} The difference between the
    delay of the first and the last arriving path,
    $\tau_\mathrm{max}=t_K-t_1$.
    \item \emph{Mean delay:} The delay corresponding to the ``centre of
    gravity'' of the PDP,

    $\tau_\mathrm{mean}=\sum^K_{k=1}\left(\rho_kt_k\right)/\sum^K_{k=1}\rho_k$.
    \item \emph{Root mean square (RMS) delay spread:} The second
    moment of PDP,

    $\tau_\mathrm{RMS}=\sqrt{\sum^K_{k=1}\left(\rho_k\tau^2_k\right)/\Sigma^K_{k=1}\rho_k-\tau^2_\mathrm{mean}}$.
\end{itemize}

We now briefly discuss some of the new insights provided by the
proposed framework. Firstly, the PDP in Fig. 2 motivated by the
classical description method of multipath channels characterizes the
relative prevalence and delay of \emph{all} the epidemic spreading
paths, whereas the existing approaches based on the theory of
complex networks only study the \emph{earliest} arrival or the
\emph{mean aggregated} behavior of infection
\cite{GBB07,GBB08,IKB17}. Secondly, the PDP captures the fundamental
relationship between the dynamics of the initial community-level
outbreak and the global, mobility-network-driven contagion. In
communications, intersymbol interference (ISI) is a form of
distortion of a signal in which one symbol interferes with
subsequent symbols. This is an unwanted phenomenon as the previous
symbols have similar effect as noise, thus making the communications
less reliable. ISI is caused by multipath propagation causing
successive symbols to ``blur'' together due to different lengths of
the paths. The RMS delay spread affects the ISI in the way that if
the symbol duration is long enough compared to the delay spread, one
can expect an equivalent ISI-free channel (i.e., a narrowband
channel). Otherwise, a wideband channel is resulted, where ISI
introduces errors in the decision device at the receiver output.
From the perspective of infection transmission, an epidemic channel
is considered ``wideband'' if the delay spread $\tau_\mathrm{RMS}$
is significant compared with the pulse duration $T_\mathrm{b}$. In
other words, if $T_\mathrm{b}>\tau_\mathrm{RMS}$, outbreaks caused
by two successive and independent patient zeroes will not interfere
with each other and the time series of all the index cases
originated from the upstream population could be successfully
recovered at the target city. In addition, techniques commonly used
in communications to alleviate ISI such as adaptive equalization and
maximum likelihood sequence estimation can also be applied to
correctly detect the encoded epidemiological states should ISI
occurs, which is beyond the scope of the current work and will be
left for future investigation.

Note that longer-scale social distancing such as air travel
restriction modifies the flux fractions over multiple epidemic
spreading paths, which in turn changes the PDP. This effort is
similar to the guiding field resulting in directed fluxes of message
carriers in the externally controllable and trackable communications
\cite{CKA15}.

\section{Numerical Examples}
In this section, we present some numerical examples to illustrate
the key principles of the proposed analytical framework. The
following parameters are used in the examples. The mean recovery
time of individuals $\beta^{-1}=5~\mathrm{days}$ ($\beta^{-1}$ is in
the range of $3$ to $5$ days for influenza-like diseases
\cite{BH13}). The initial growth time
$\alpha^{-1}=1.25~\mathrm{days}$, which results in the basic
reproduction ratio of $\alpha/\beta=4$ (reproductive number
estimates are in the range of $1.46$ to $4.48$ for the epidemic
influenza \cite{SL84}). The final growth rate is $0$ after
successfully implementing dynamic mass vaccination and
community-level social distancing. Four different rates of decline
are considered, $q^{-1}=8,9,10,11~\mathrm{days}$. The threshold
$\epsilon$ defining the duration of the epidemic pulse is set to be
$1$.

In terms of the underlying traffic, it is assumed that the number of
paths is $10$ and the maximum number of hopping for each path is
$6$. The maximum traffic flux is $0.01$ (i.e., the fraction of
people traveling between two successive nodes per unit time is
$1\%$; the absolute sizes of the populations are not required),
consistent with the typical scenario in a major transport hub in the
global mobility network \cite{BH13}. \begin{figure} [!htp]
\begin{center}
\epsfig{file=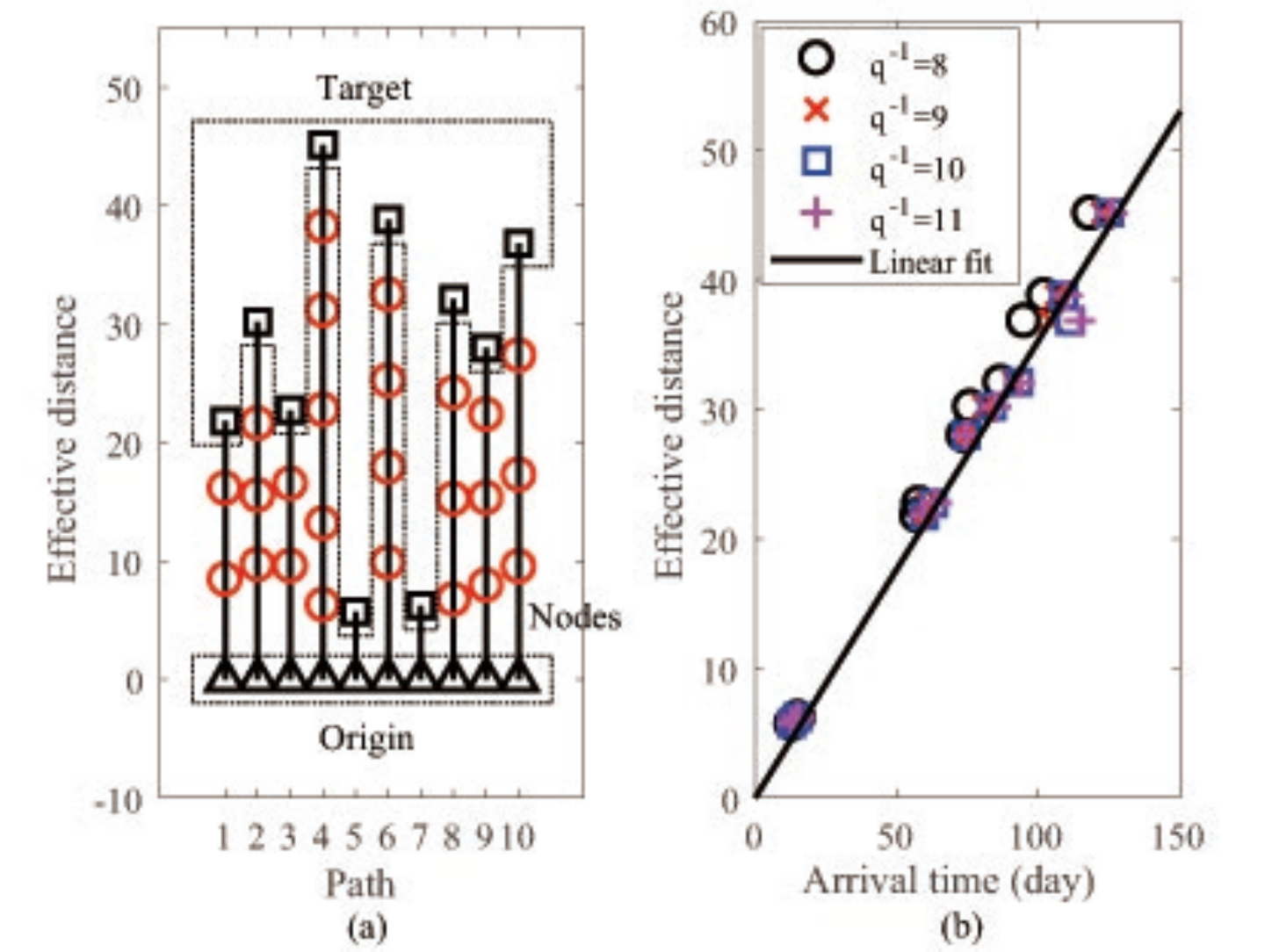,width=0.8\linewidth}
\end{center}
    \caption{(a) Multipath structure of the mobility network for epidemic spreading
    from the origin subpopulation (``$\triangle$'') to the target subpopulation (``$\square$'')
    \emph{via} intermediate nodes (``$\circ$''), and (b) the relationship between the
    overall effective distance and the arrival time for different values of the pulse shaping parameter $q$.}
    \label{fig:Fig3}
\end{figure}
\begin{figure} [!htp]
\begin{center}
\epsfig{file=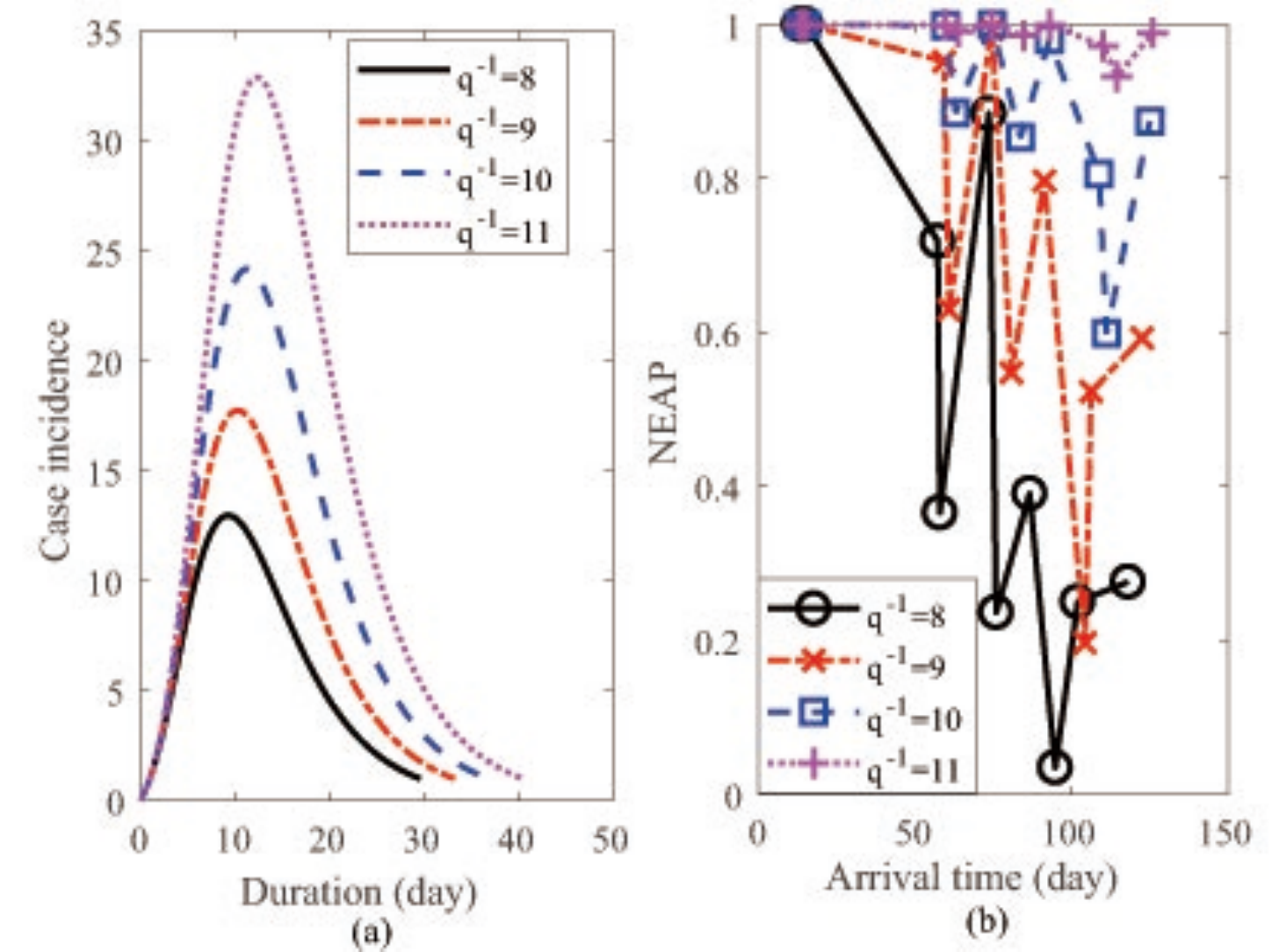,width=0.8\linewidth}
\end{center}
    \caption{(a) The epidemic pulses and (b) the corresponding PDPs for different values of the pulse shaping parameter $q$.}
    \label{fig:Fig4}
\end{figure}
\begin{figure} [!htp]
\begin{center}
\epsfig{file=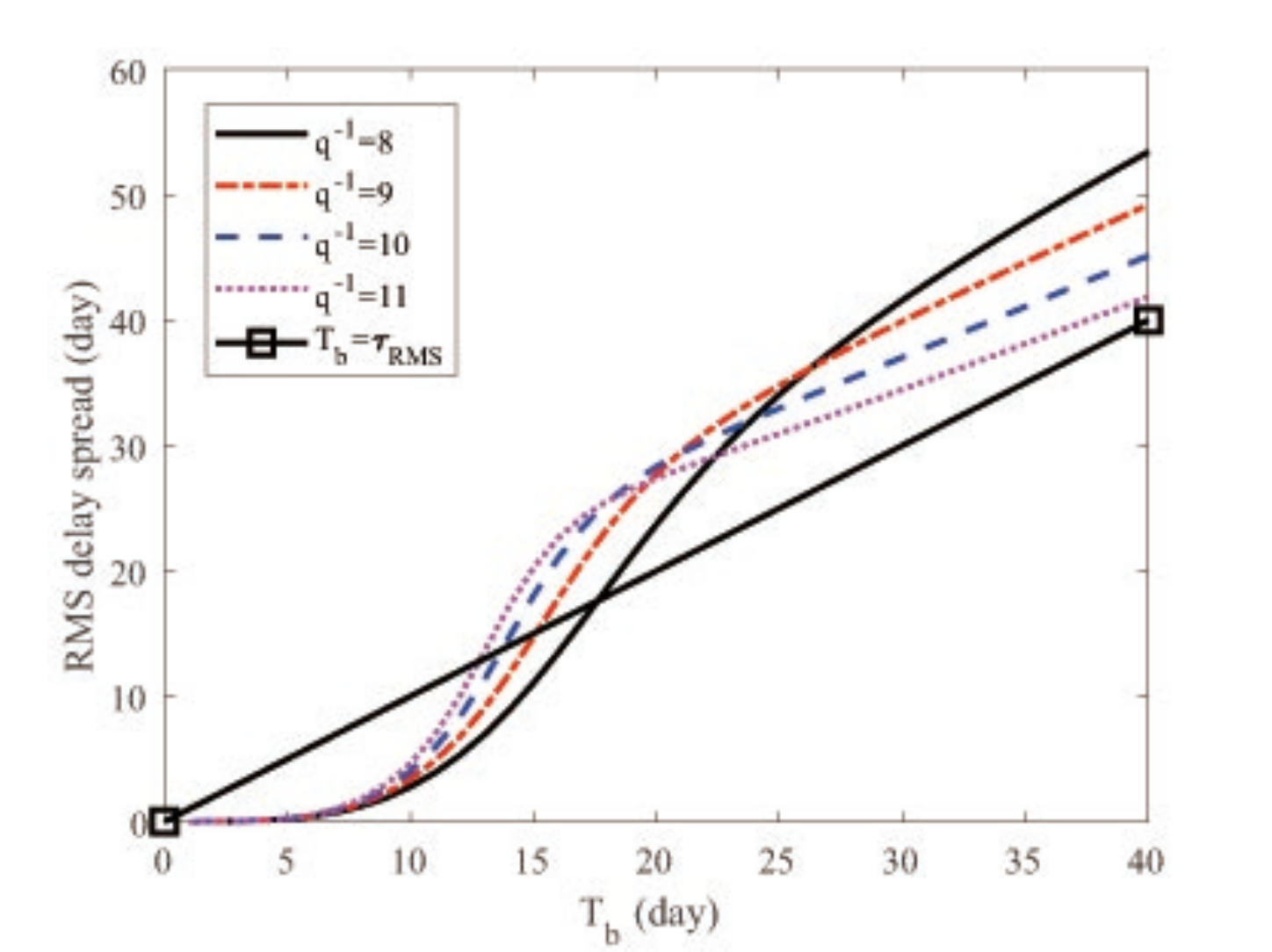,width=0.8\linewidth}
\end{center}
    \caption{The relationship between the RMS delay spread $\tau_\mathrm{RMS}$ and the pulse duration $T_\mathrm{b}$
    for different values of the pulse shaping parameter $q$. The boundary of the narrowband and wideband regimes is
    defined by $T_\mathrm{b}=\tau_\mathrm{RMS}$.}
    \label{fig:Fig5}
\end{figure}
The minimum traffic flux is $0.0001$. The effective distance for
each hopping is supposed to be uniformly distributed between
$1-\ln(0.01)$ and $1-\ln(0.0001)$. The focus on the parameters is
not essential, as the proposed framework is also valid for other
types of mobility dynamics.

Fig. 3(a) depicts the multipath structure of a simulated mobility
network for epidemic spreading from the origin to the target
subpopulation \emph{via} multiple intermediate nodes. Note that all
the triangles in the figure refer to the same origin and all the
squares refer to the same target. For each path, the locations of
intermediate nodes are indicated in terms of their effective
distances from the origin. As the epidemic spreading is analyzed
from the communications perspective rather than the complex network,
the multipath representation explicitly showing effective distances
for each path is used in Fig. 3(a), which is easier to understand
than the traditional network topology. Fig. 3(b) depicts the
relationship between the overall effective distance for each path,
which is obtained by adding up effective distances between
successive nodes along that path, and the arrival time for the
network in Fig. 3(a). A strong correlation between effective
distance and arrival time is evident. Furthermore, the relationship
is linear, meaning that the effective speed of the wavefront is a
well-defined constant. This phenomenon is a consistent extension of
the observations made in \cite{BH13,IKB17}. It is worth noting that
a larger pulse shaping parameter $q$ (see also Fig. 4(a)) results in
a slightly bigger effective speed even for the same underlying
network. This is in contrast to the communications channel where the
speed of message carrier is a property of propagation medium but not
transmitted signal.

Fig. 4(a) illustrates the epidemic pulses transmitted from the
origin for different values of $q$, which corresponds to the case
incidence in the origin subpopulation. As could be expected, slower
rate of decline of the disease growth results in broader pulses. The
corresponding PDPs are shown in Fig. 4(b) given the same mobility
network generated. We can make the following important observations.
Firstly, overall larger arrival time leads to lower prevalence due
to reduced probability in disease transmission. Nevertheless, the
relationship is not monotonic as multiple hops along a path
potentially increase the chance of epidemic spreading due to reduced
distances between neighboring nodes, though the overall arrival time
increases. Secondly, as the rate of decline of the outbreak growth
decreases, the total access delay increases and the weights of all
paths approach $1$ (i.e., the prevalence becomes less sensitive to
the arrival time).

Fig. 5 demonstrates the relationship between the RMS delay spread
$\tau_\mathrm{RMS}$ and the pulse duration $T_\mathrm{b}$ for
different values of $q$. Suppose that $T_\mathrm{b}$ needs to be
greater than $\tau_\mathrm{RMS}$ to ensure a narrowband epidemic
channel such that the pulse shape of $C(t)$ would be reserved at the
target node for reliable analysis of early contagion dynamics. As
shown in Fig. 5, a crossover phenomenon can be observed as $q$
decreases where smaller $q$ results in larger $\tau_\mathrm{RMS}$ in
the regime of small $T_\mathrm{b}$ and an opposite trend occurs for
large $T_\mathrm{b}$. Furthermore, the epidemic channel appears to
be narrowband for smaller $T_\mathrm{b}$, and gradually becomes
wideband as $T_\mathrm{b}$ increases. However, the trend is
dependent on the specified boundary of the narrowband and wideband
regimes in the $\tau_\mathrm{RMS}-T_\mathrm{b}$ plane as well as the
underlying multipath structure of the mobility network generated.

\section{Conclusion}
In summary, the analysis of air-traffic-mediated epidemic spreading
and mitigation strategies in the communications-inspired framework
enables researchers to understand complex contagion dynamics by
applying the deep-rooted communications theories. We have presented
the quantity of PDP to describe the relative influence of multiple
disease transmission paths, and introduced the metric of RMS delay
spread to measure the frequency selectivity of an epidemic channel
(``narrowband'' vs. ``wideband'').

The method is a useful starting point for more detailed
investigations. Firstly, the framework would motivate acquisition of
real spatiotemporal epidemic data (PDPs at various target locations)
that reflect the multipath nature of infection spreading. Secondly,
the approach allows utilizing a wide range of signal detection
algorithms and performance measures traditionally only used in the
communications field (e.g., equalization, bit error rate) for
analysis of global disease dynamics and social/medical
interventions. Finally, the communications-inspired model may be
extended to other contagion phenomena, such as human-mediated spread
of violence or rumors.

\section*{Acknowledgement}
This work is supported by the Guangdong Natural Science Funds under
Grant 2016A030313640.

\bibliographystyle{IEEEbib}
\bibliography{IEEEbib}
\let\nofiles\relax

\end{document}